\documentclass[aps,prl,twocolumn,amsmath,showpacs,amssymb]{revtex4}
\usepackage{graphicx}	
\usepackage{epsf}		
\usepackage{natbib}
\include{epsfig}
\usepackage{bbm}
\usepackage{mathrsfs}
\usepackage{amsfonts}
\usepackage{color}
\newcommand{\be}{\begin{equation}}
\newcommand{\ee}{\end{equation}}
\newcommand{\bea}{\begin{eqnarray}}
\newcommand{\eea}{\end{eqnarray}}
\begin{document}
\title{The Poisson-Boltzmann Theory for The Two-Plates Problem: Some Exact Results}
\author{Xiangjun Xing}
\address{Institute of Natural Sciences and Department of Physics, 
Shanghai Jiao Tong University,
Shanghai, 200240 China}

\date{\today} 
\begin{abstract} 
The general solution to the nonlinear Poisson-Boltzmann equation for two parallel charged plates, either inside a symmetric elecrolyte, or inside a $2\mbox{q:-q}$ asymmetric electrolyte, is found in terms of Weierstrass elliptic functions.  From this we derive some exact asymptotic results for the interaction between charged plates, as well as the exact form of the renormalized surface charge density.   
\end{abstract}
\pacs{82.70.Dd, 83.80.Hj, 82.45.Gj, 52.25.Kn}

\maketitle
\begin{figure}
\begin{center}
\includegraphics[width=7cm]{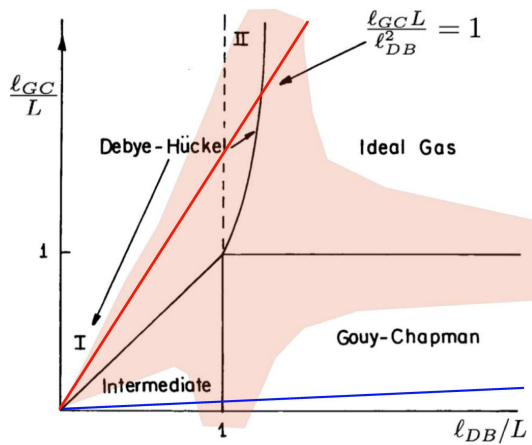}
\caption{The parameter space for the two plate problem.   There are four regimes where the asymptotic are well understood.  Our solution in terms of Weierstrass elliptic functions reduces to the known asymptotic results in the corresponding regimes, and works in the cross-over regions shown shaded in the figure as well. Part of this plot was taken from the review by Andelman \cite{Andelman-PBequation}.  The blue line: two strongly charged plates with a variable distance.  The red line: two weakly charged plates with a variable distance. }
\label{four-regimes}
\end{center}
\vspace{-5mm}
\end{figure}

A charged object inside an electrolyte attracts ions of opposite charge and repels ions of like charge.  The total electrostatic potential, due to both the external charges and the electrolyte, is exponential damped as a function of distance from the charged object.  This screening phenomenon, first studied by Debye and Huckel (DH) \cite{Debye-Huckel}, is the most essential property of ionized matter.  The Poisson-Boltzmann (PB) equation associated with the screened potential is one of the most important equations in science and engineering.  Despite of this, our understanding of PB is very limited, largely due to its nonlinear nature.  Analytic solutions are known only for a few cases, i.e.  1) one plate problem \cite{Verwey-Overbeek, Andrietti19761121}; and 2) one cylinder problem \cite{Tracy-Widom-Physica-1997}, both in $\mbox{q:-q}$ symmetric, and in $2\mbox{q:-q}$ asymmetric electrolytes.  The two plate problem is the simplest toy problem for effective interactions between charged objects, and therefore has received constant attention ever since Debye-H\"{uckel}.   Its solution in symmetric electrolytes was expressed in terms of elliptic integrals in the classical monograph by Verwey and Overbeek \cite{Verwey-Overbeek}.  From this the asymptotic behaviors can be derrived in four different regimes, shown in Fig.~\ref{four-regimes}.  For a detailed account, see the review by Andelman \cite{Andelman-PBequation}.  The behaviors in cross-over regions however are not understood.  Moreover, the two-plate problem in asymmetric electrolytes has not been touched.  In this work, we outline a number of new exact results for the two plate problem both in $\mbox{q:-q}$ and in $2\mbox{q:-q}$ electrolytes.   Detailed discussion will be published elsewhere \cite{PB-two-plates-xing}. 

Let us start with the Poisson-Boltzmann equations for the reduced, dimensionless potential $\Psi = \beta q \varphi$: 
\begin{subequations}
\label{saddle}
\bea
 - \Delta \Psi +  \sinh \Psi &=& 0, 
 \,\,\, \quad \quad \mbox{q:-q}; 
 \label{saddle-sym}
 \\ 
  -  \Delta \Psi +  e^{\Psi} /3
   - e^{-2 \Psi} /3&=& 0, 
   \quad \quad 2\mbox{q:-q}.  
\label{saddle-asym}
\eea
\end{subequations}
where Eq.~(\ref{saddle-sym}) applies to $\mbox{q:-q}$ symmetric electrolytes, while Eq.~(\ref{saddle-asym}) to $2\mbox{q:-q}$ asymmetric electrolytes.  $\beta$ is the Boltzmann factor, $q= 1.6\times 10^{-19}C$ the charge of an electron.  The real space coordinates are measured in unit of the Debye length:
\be
\ell_{DB}= \kappa^{-1} = 
\left\{
\begin{array}{ll}
 \sqrt{\epsilon/2 \beta q^2 n},
 &  \,\,\, \quad \mbox{q:-q} ;\vspace{3mm}\\
\sqrt{\epsilon/3 \beta q^2 n},
 & \quad \mbox{2q:-q}.
\end{array}
\right.
\label{Debye-length-def}
\ee  
For the problem of two charged plates of infinite size, $\Psi$ only depends on one coordinate $z$. The following first integrals of Eqs.~(\ref{saddle}) are easily obtained:
\bea 
\begin{array}{ll}
{\alpha} =- \frac{1}{2} (\partial_z \Psi)^2 +  \cosh \Psi (z), 
& \,\,\,  \quad \mbox{q:-q} ; 
\vspace{3mm}\\
\alpha = - \frac{1}{2} (\partial_z \Psi)^2 
+ \frac{1}{3}e^{\Psi} + \frac{1}{6} e^{-2 \Psi}, 
& \quad \mbox{2q:-q}. 
\end{array} 
\label{first-integral}
\eea
In the middle of two plates, $\partial_z \Psi (MP) = 0$, therefore $\alpha$ is related to $\Psi(MP)$ via
\bea 
\begin{array}{ll}
{\alpha} = \cosh \Psi (MP), 
& \,\,\,  \quad \mbox{q:-q} ; 
\vspace{3mm}\\
\alpha = 
\frac{1}{3}e^{\Psi(MP)} + \frac{1}{6} e^{-2 \Psi(MP)}, 
& \quad \mbox{2q:-q}. 
\end{array} 
\label{first-integral-0}
\eea
If two plates are separated by infinite distance, $\Psi(MP) = 0$, therefore $\alpha$ reduces to $\alpha _0 = 1$ \, (\mbox{q:-q}) and $\alpha _0 =1/2$ \,(2\mbox{q:-q}) respectively. 
The net interaction between two plates is proportional to $\delta \alpha = \alpha - \alpha_0$:  
\be
P_{\rm net} 
= \frac{T\,\delta \alpha}{ 4 \pi \ell_{DB}^2 \lambda_{Bj}},  
\label{P_net}
\ee
with $\lambda_{Bj} = q^2/4 \pi \epsilon T$ the Bjerrum length.  We shall therefore call $\delta \alpha$ the reduced, dimensionless interaction.  Note $P_{\rm net}$ vanishes as the separation becomes large. 
For detailed derivations of these results, see reference \cite{PB-two-plates-xing}.

\begin{figure}
\begin{center}
\includegraphics[width=6cm]{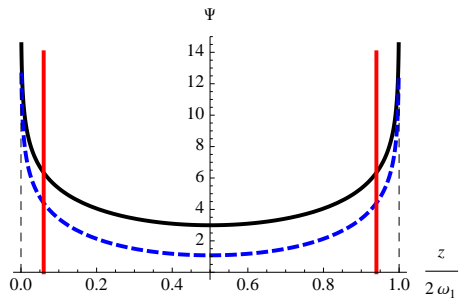}
\caption{
The dimensionless potential  Eqs.~(\ref{Psi-p}), with $z$ a real variable.  Solid: $\alpha =10$ in $\mbox{q:-q}$ symmetric electrolyte. Dashed: $\alpha = 1$ in $2\mbox{q:-q}$ asymmetric electrolyte.  $z = 0, 2 \omega_1$ are logarithmic singularities.  Two dashed vertical lines are two maximally separated plates with infinite surface charge density, for given $\alpha$.  Two solid red lines are two plates with finite surface charge density that gives the same pressure. }
\label{wp-1}
\label{potential-plot}
\end{center}
\vspace{-5mm}
\end{figure}

Now introducing a new function $\wp(z)$ through 
\be
\begin{array}{ll}
4 \wp(z) + 2 \alpha / 3 = e^{\Psi(z)} , 
&
\,\,\, \quad \mbox{q:-q}; 
\vspace{2mm}\\
6 \wp(z) + \alpha = e^{\Psi(z)} ,
&  \quad \mbox{2q:-q},
\end{array}
\label{y-Psi}
\ee
the first integrals Eqs.~(\ref{first-integral}) are transformed into 
\be
(\wp')^2 = 4\, \wp^3 - g_2 \, \wp - g_3,
\label{Elliptic-ODE}
\ee
where 
\bea
( g_2, g_3 )=  
\left\{ \begin{array}{ll}
( \frac{\alpha ^2}{3} - \frac{1}{4} ,
  \frac{\alpha ^3}{27} - \frac{\alpha}{24}), 
  & \,\,\, \quad  \mbox{q:-q} ;
  \vspace{3mm}\\
  ( \frac{\alpha ^2}{3} ,
  \frac{\alpha ^3}{27} - \frac{1}{108} ), 
  & \quad \mbox{2q:-q}. 
  \end{array} \right.  
\label{g-A}
\eea
Eq.~(\ref{Elliptic-ODE}) is precisely the differential equation satisfied by the doubly periodic {\em Weierstrass elliptic function} $\wp(z;g_2,g_3)$, with $g_2$ and $g_3$ two of its {\em invariants} given by Eqs.~(\ref{g-A}).  $\wp(z)$ can be represented as a double series in $z$:
\be
\wp(z) = \frac{1}{z^2} + {\sum_{m,n}} ' \left[
\frac{1}{ (z - 2 m \omega_1 - 2 n \omega_2 )^2 } 
- \frac{1}{( 2 m \omega_1 + 2 n \omega_2)^2} 
\right],
\label{P-series}
\ee
where the prime in the summation means exclusion of the term with $m = n = 0$, and $z$ is generally treated as a {\em complex variable}.  $\wp(z)$ is a meromorphic function of $z$ with an infinite number of second order poles $z_{m,n} = 2 m \omega_1 + 2 n \omega_2$, which form a 2D rectangle lattice for $\alpha > \alpha_0$.  We shall choose $\omega_1$ to be real and $\omega_2$ to be purely imaginary.  One can easily check using Eq.~(\ref{P-series}) that $\wp(z)$ is periodic with respect to two complex {\em periods} $2 \omega_1$ and $2 \omega_2$: $\wp(z) = \wp(z+ 2 \omega_1) = \wp(z + 2 \omega_2)$. There is one-to-one relation between the periods $2 \omega_1,2\omega_2$ and two invariants $g_2,g_3$, which can be easily computed using Wolfram Mathematica 7.  In our case, they are all uniquely determined by one parameter $\alpha$, see Eq.~(\ref{g-A}). The readers are referred to the classic monograph by Whittaker and Waston \cite{Whittaker-Watson} for relevant details.

Inverting Eqs.~(\ref{y-Psi}), we find the potential $\Psi$ as 
\be
\Psi(z) = \log \left[ 
\wp \left(z; g_2,g_3 \right) + \frac{\alpha}{6}
\right] +\left(
\begin{array}{c}
 \log 4\\ \log 6\end{array}\right),    \label{Psi-p}
\ee
where the upper/lower case is for $\mbox{q:-q}$ and $2\mbox{q:-q}$ electrolytes respectively. 
Eq.~(\ref{Psi-p}) is always positive and has logarithmic singularities at $z = 0, 2\omega_1$, as illustrated in Fig.~\ref{potential-plot}.  It therefore describes the potential between two {\em positively} charged plates.  Since the potential between plates must be a smooth function, these two singularities $0, 2 \omega_1$ must be {\em outside two plates}.  Hence for a given interaction $\alpha$, there is a {\em maximum separation} between two plates, which is precisely the real period $2 \omega_1$.  {\em The surface charge density yielding this interaction $\alpha$ at this maximal separation is infinity}, see the dashed vertical lines in Fig.~\ref{potential-plot}.  This constitutes a direct proof that the interaction between two plates remains finite as the surface charge density scales up to infinity, a property 
usually called ``charge renormalization'' \cite{charge-RG-Alexander-JCP-1984}.  It has been repeatedly observed in numerical studies of PB equation.

Taking the limit $\alpha \rightarrow \alpha_0$, we find that two invariants Eq.~(\ref{g-A}) reduce to $(1/12, -1/216)$ for both cases.  Therefore Eq.~(\ref{Psi-p}) reduces to 
\be
\Psi(z) = \log \left[ 
\wp \left(z; \frac{1}{12}, - \frac{1}{216} \right)
 + \left(
\begin{array}{c}
 1/6\\ 1/12\end{array}\right)
\right] +\left(
\begin{array}{c}
 \log 4\\ \log 6\end{array}\right).   \label{Psi-p-0}
\ee
They can be rewritten into
\begin{subequations}
\bea
\Psi^{\pm}_{\mbox{q:-q}}(z) &=&  
 2 \log \frac{1\pm e^{- z }}
{1\mp e^{- z  }},  
\label{potential-one-plate}
\\
\Psi_{2\mbox{q:-q}}^{\pm}(z) 
&=& \log \frac{1 \pm 4 \, e^{-z}
 + e^{-2z}}{(1 \mp e^{-z})^2}. 
\label{potential-one-plate-1}
\eea
\label{potential-one}
\end{subequations}
The result Eq.~(\ref{potential-one-plate}) for $\mbox{q:-q}$ electrolyte has been known since the time of Verwey and Overbeek \cite{Verwey-Overbeek}.  The result Eq.~(\ref{potential-one-plate-1}) for $2\mbox{q:-q}$ electrolyte was first discovered by Andrietti {\it et al} in 1976 \cite{Andrietti19761121}, but has remained largely unknown since then.  Our analysis makes it clear that these solutions are related to one and the same elliptic function, with its real period approaching infinity.  

The general solutions of the one plate problem are obviously Eqs.~(\ref{potential-one}) translated by an arbitrary constant $z^*$, whose value is to be determined by the Neumann boundary condition.  For our purpose, however, it is more convenient to fix the solution to be Eqs.~(\ref{potential-one}), and adjust the position of the plate $z_0$ to satisfy the boundary condition: 
\be
\partial \Psi/\partial z|_{z_0} = - \eta. 
\label{BC-eta}
\ee
where
$\eta ={\beta q \sigma}/{ \epsilon \kappa} 
= {\ell_{DB}}/{\ell_{GC}}$ is the dimensionless surface charge density, with 
$\ell_{GC} =  \epsilon/ q \beta \sigma$ the Gouy-Chapman length.  


\begin{figure}
\begin{center}
\includegraphics[width=6cm]{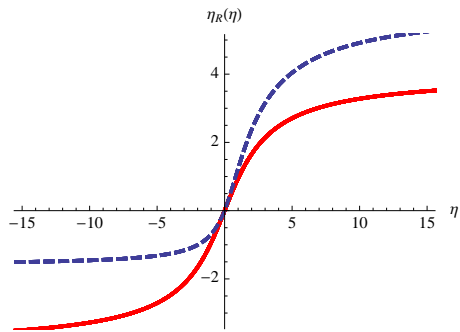}
\caption{ The renormalized surface charge density $\eta_R$  as a function of the bare surface charge density $\eta$ for a charged plate as given by Eqs.~(\ref{relation-eta-eta_R}).  Red solid line: $\mbox{q:-q}$.  $\eta_R$ saturates at $4$ in both directions.  Blue dashed line: $2\mbox{q:-q}$. $\eta_R$ saturates at $6$ and $6(2-\sqrt{3})$ in the positive and negative sides respectively.  } 
\label{potential-oneplate}
\label{effectiveQ-oneplate}
\end{center}
\vspace{-5mm}
\end{figure}


The far field asymptotics of Eqs.~(\ref{potential-one}) are given by 
$\pm 4\,e^{-z}, \pm 6 e^{-z}$ respectively.  On the other hand, in the linear PB theory, a plate with surface charge density $\eta_R$ at the same location $z_0$ produces at $z$ a potential 
\be
\Psi^{\rm linear} (z) = \eta_R \, e^{-z + z_0},  
\label{solution-linear-BP}
\ee
We identify the far-field asymptotics of the nonlinear theory with the linear theory Eq.~(\ref{solution-linear-BP}), and define the renormalized (or effective) charge density $\eta_R$.   Using Eq.~(\ref{BC-eta}) to eliminate $z_0$ in favor of $\eta$, we find $\eta_R (\eta)$ as a function of the bare surface charge density $\eta$: 
\bea
\begin{array}{ll} 
\eta_R(\eta) =\frac{2 \eta}
{ 1+ \sqrt{1+(\eta/2 )^2} },
& \,\,\, \quad \mbox{q:-q} , 
\vspace{3mm}\\
\frac{36 \eta _R \left(\eta
   _R+6\right)}{\left(6-\eta _R\right)
   \left(\eta _R^2 + 24 \eta_R +36\right)} 
   = \eta,
& \quad 2\mbox{q:-q},
 \end{array}
 \label{relation-eta-eta_R}  
\eea
which are illustrated in Fig.~\ref{effectiveQ-oneplate}.  In the weakly charged limit $\eta \ll 1$, $\eta_R \rightarrow \eta $; in the strongly charged limit $\eta \gg 1$, $\eta_R$ saturates at $\eta_R( \pm \infty) = \pm 4$ for symmetric electrolytes and $\eta_R(\infty) = 6, \eta_R(-\infty) = -6 (2- \sqrt{3})$ for $2\mbox{q:-q}$ asymmetric electrolytes.  The practice of using linear theory with an renormalized, i.e. effective, surface charge density in the far field, is usually called {\em charge renormalization} following the seminal work by Alexander {\it et al}.  Here the most striking property is that $\eta_R$ saturates at a finite value in the strongly charged limit.  {\em It shows that electrolytes are able to screen infinitely charged objects within finite distance.}  This should be regarded as one fundamental property of the Poisson-Boltzmann theory.  

If two plates are widely separated, $L \gg \ell_{DB}$, the potential between is approximately the sum of that of two isolated plates.  The potential in the middle is
\be
\Psi(MP) =  2 \eta_R(\eta) \, e^{-L/2\ell_{DB}},
\ee
up to corrections smaller by powers of $e^{-L/\ell_{DB}}$.  Note that we have restored the physical unit for length here.  Using Eqs.~(\ref{first-integral-0}) and Eq.~(\ref{P_net}), the net interaction between two plates is 
\be
P_{\rm net} = 
\frac{T \eta_R(\eta)^2}{2 \pi \ell_{DB}^2 \lambda_{Bj}}
 e^{-L/\ell_{DB}}, 
 \quad \quad  L \gg \ell_{DB}, 
 \ee
 where $\eta_R$ is determined by Eqs.~(\ref{relation-eta-eta_R}).   This result applies to the whole region to the left of the vertical line $\ell_{DB}/L = 1$ in Fig.~\ref{four-regimes}, which includes the intermediate regime as well as a major part of the Debye-H\"{u}ckel regime. 
   
Now consider two infinitely charged plates separated by a small distance $L \ll \ell_{DB}$.   the potential between plates is high, so that the co-ion density is negligibly small. The corresponding term is exponentially small and can be dropped from the PB equation.  The resulting equation can be integrated easily. After some simple calculations, one obtains the pressure, in SI unit, as
\be
P =  \left\{ \begin{array}{ll}
 \frac{2 \pi^2\epsilon T^2}{q^2 L^2}, 
&  \pm \infty \,\, \mbox{plates} \,\,\,\,\,\mbox{q:-q} 
\vspace{2mm}\\
 \frac{2 \pi^2\epsilon T^2}{q^2 L^2}, 
&  +\infty \,\, \mbox{plates} \,\,\mbox{2q:-q}
\vspace{2mm}\\
 \frac{\pi^2\epsilon T^2}{2 q^2 L^2}, 
& - \infty \,\, \mbox{plates} \,\,2\mbox{q:-q}
\end{array}
\right. .
\ee 
This is called the {\em Gouy-Chapman regime} \cite{Andelman-PBequation}.

Consider the case of symmetric electrolytes.  Assume $L \ll \ell_{DB}, \ell_{GC}$
The potential Eq.~(\ref{Psi-p}) between plates is small and can be expanded into Taylor series around the mid-point up to quadratic order of $\delta z = z - \omega_1$: 
\be
\Psi(z) = \Psi(\omega_1) + \frac{1}{2} \Psi''(\omega_1) \delta z^2
+ O(\delta z^4). 
\ee
Using the properties of Weierstrass elliptic functions:, we find $\Psi''(\omega_1) = \sqrt{\alpha^2 - 1}$.  For details, see reference \cite{PB-two-plates-xing}.  The Neumann boundary condition Eq.~(\ref{BC-eta}) evaluated at $\delta z = L/2$  then gives $\alpha$ as a function of $\eta, L$.  Substituting this back into Eq.~(\ref{P_net}), we find: 
\be
P_{\rm net} = \frac{T}{4 \pi \ell_{DB}^2 \lambda_{Bj}} 
\left( \sqrt{1 + \left( \frac{2 \ell_{DB}^2 } {\ell_{GC} L } \right) ^2} -1 \right).
\ee 
This result is applicable in the top right quarter of Fig.~\ref{four-regimes}, which includes the ideal gas regime, and part of the DH regime.  Similar analyses can also be carried out for the cases of asymmetric electrolytes.  The results however are rather complicated, and not particularly illuminating.  We therefore do not present them here.

We shall not discuss the asymptotic behaviors in DH regime, as they are the same to {\em all} electrolytes and have been presented in various literatures. See, for example, the review article by Andelman \cite{Andelman-PBequation}.  

\begin{figure}
\begin{center}
\includegraphics[width=5.5cm]{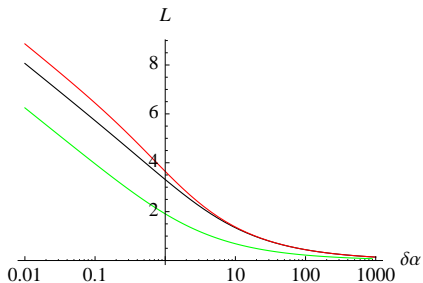}
\caption{ Log-linear plots of $L$ v.s. $\delta \alpha$ for two infinitely charged plates.  
Black: $\mbox{q:-q}$ symmetric electrolyte;  
Red: $\eta = +\infty$, $2\mbox{q:-q}$ electrolyte, see Eq.~(\ref{alpha-L-max}).
Green: $\eta = - \infty$, $2\mbox{q:-q}$ electrolyte, see Eq.~(\ref{alpha-L-max-asym-2}).  
}
\vspace{-5mm}
\label{period-A-plot}
\end{center}
\end{figure}

Now consider two infinitely charged plates.  According to Fig.~(\ref{potential-plot}), the separation between plates is just the real period $2 \omega_1\equiv 2 \omega_1 (\alpha) = \Upsilon_{\rm WHP}
\left( g_2(\alpha),g_3(\alpha)\right)$,
with $g_k(\alpha)$ given by Eqs.~(\ref{g-A}).  Here the subscript ``WHP'' stands for ``Weierstrass Half period''. The function $\Upsilon_{\rm WHP}$ can be conveniently computed using Wolfram Mathematica 7.  The interaction between two $\eta = \pm \infty$ plates in symmetric electrolytes, or two $\eta = + \infty$ plates in $2\mbox{q:-q}$ electrolytes, are given by 
\be
L = 2\, \Upsilon_{\rm WHP}
\left( g_2(\alpha),g_3(\alpha)\right). 
\label{alpha-L-max}
\ee

The case of negatively charged plates in 2q:-q electrolytes is slightly more complicated.  Here we refer the readers to reference \cite{PB-two-plates-xing} for details, and present the result directly.  The relation between $L$ and $\alpha$ for a pair of negative infinitely charged plates is given by 
\bea
L 
=  2\,\Upsilon_{\rm WHP} \left( 
\frac{\alpha ^2}{3} ,
\frac{\alpha ^3}{27} - \frac{1}{108} 
\right)
- 2 \zeta_1(\alpha),
\nonumber \\  - \infty \,\, \mbox{plates} \,\,
 2\mbox{q:-q}, 
\label{alpha-L-max-asym-2}
\eea
where $\zeta_1$ is the smaller of the two zeros of the function $\wp\left(z + \omega_2; \frac{\alpha ^2}{3}, \frac{\alpha ^3}{27} - \frac{1}{108} \right) + \alpha/6 $ in the period $(0,2 \omega_1)$.  The results Eqs.~(\ref{alpha-L-max}, \ref{alpha-L-max-asym-2}) are plotted in Fig.~\ref{period-A-plot}.  

For high but finite surface charge density, we have 
\be
L = 2 \omega_1(\alpha) - 2 z_0,
\label{z0A-1}
\ee
where $z_0$ is the position of the left plate.  The Neumann boundary condition in Eq.~(\ref{BC-eta}) can be used to solve for $z_0$ in terms of $\eta$.  Plugging the results back into Eq.~(\ref{z0A-1}) we find  
\begin{subequations}
\bea
&& 2 \,\Upsilon_{\rm WHP} \left( 
\frac{\alpha ^2}{3} - \frac{1}{4},
\frac{\alpha ^3}{27} - \frac{\alpha}{24}
\right)
= L + \frac{4}{\eta},  
\quad \quad\quad \quad \quad \mbox{q:-q} 
\nonumber
\\
&& 2 \,\Upsilon_{\rm WHP} \left( 
\frac{\alpha ^2}{3},
\frac{\alpha ^3}{27} - \frac{1}{108}
\right)
= L + \frac{4}{\eta},  
\quad \quad \quad \quad \quad
+ \,\mbox{2q:-q}
\nonumber
\\
&& 2 \,\Upsilon_{\rm WHP} \left( 
\frac{\alpha ^2}{3},
\frac{\alpha ^3}{27} - \frac{1}{108}
\right)
= L + 2 \zeta_1(\alpha) - \frac{2}{\eta},  
\quad 
- \,\mbox{2q:-q}
\nonumber
\eea
\label{eqn-alpha-0}
\end{subequations}
which holds in the whole strongly charged regime where the Gouy-Chapman length $\ell_{GC}$ is much shorter than the Debye length $\ell_{DB}$.  This includes many examples in biological physics.  For a given surface charge density $\eta$, the $\alpha-L$ curve corresponding to Eq.~(\ref{eqn-alpha-0}) can be obtained by a rigid shift of the saturation curve corresponding to Eq.~(\ref{alpha-L-max},\ref{alpha-L-max-asym-2}) along the $L$ axis.



\begin{figure}
\begin{center}
\includegraphics[width=6cm]{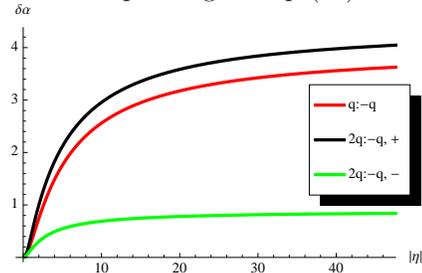}
\caption{The net  pressure $(\alpha - \alpha_0)$ as a function of the surface charge density $|\eta|$ for given plate separation $L = 2$.  All quantities are dimensionless.  Red: charged plates in a symmetric electrolyte.  Black: positively charged plates in a $2\mbox{q:-q}$ asymmetric electrolyte.  Green: negatively charged plates in a $2\mbox{q:-q}$ electrolyte.  Note how multi-valence counter-ions can dramatically reduce the pressure between two plates.  }
\label{alpha-eta}
\end{center}
\vspace{-5mm}
\end{figure}

Finally for the general case of arbitrary surface charge density, we can numerically solve for $\alpha$ with given $L, \eta$.  
For example, we plot for given value of $L = 2 \ell_{DB}$, the  pressure $\alpha$ as a function of the surface charge density $\eta$ in Fig.~\ref{alpha-eta} for all three cases.  Note how the  pressure saturates as the surface charge density $\eta$ increases.  One can also see explicitly how divalence counter-ions dramatically reduces the repulsion between the plates, comparing with the case of mono-valence counter-ions.  With slightly more efforts, we can also plot $\alpha $ as a function of separation $L$ for given surface charge density $\eta$.  We shall however not discuss this in detail here.  

The author thanks Leo Radzihovsky,  Andy Lau, Hongru Ma, Erik Luijten, Michael Brenner, and Anatoly Kolomeisky for interesting discussions on electrolytes.

\bibliography{/Users/xxing/research/reference-all}

\end{document}